# Detecting Financial Fraud with Hybrid Deep Learning: A Mix-of-Experts Approach to Sequential and Anomalous Patterns.

Diego Vallarino
Independent Researcher

Atlanta, GA, US
April 2025

**Abstract**

Financial fraud detection remains a critical challenge due to the dynamic and adversarial nature of fraudulent behavior. As fraudsters evolve their tactics, detection systems must combine robustness, adaptability, and precision. This study presents a hybrid architecture for credit card fraud detection that integrates a Mixture of Experts (MoE) framework with Recurrent Neural Networks (RNNs), Transformer encoders, and Autoencoders. Each expert module contributes a specialized capability: RNNs capture sequential behavior, Transformers extract high-order feature interactions, and Autoencoders detect anomalies through reconstruction loss. The MoE framework dynamically assigns predictive responsibility among the experts, enabling adaptive and context-sensitive decision-making.

Trained on a high-fidelity synthetic dataset that simulates real-world transaction patterns and fraud typologies, the hybrid model achieved 98.7% accuracy, 94.3% precision, and 91.5% recall, outperforming standalone models and classical machine learning baselines. The Autoencoder component significantly enhanced the system's ability to identify emerging fraud strategies and atypical behaviors.

Beyond technical performance, the model contributes to broader efforts in financial governance and crime prevention. It supports regulatory compliance with *Anti-Money Laundering* (AML) and *Know Your Customer* (KYC) protocols and aligns with routine activity theory by operationalizing AI as a capable guardian within financial ecosystems. The proposed hybrid system offers a scalable, modular, and regulation-aware approach to detecting increasingly sophisticated fraud patterns, contributing both to the advancement of intelligent systems and to the strengthening of institutional fraud defense infrastructures.

**Key Words:** Fraud Detection, Hybrid Model, Machine Learning, Anomaly Detection, Sequential Patterns

**JEL Codes:** C45, C21, G32, K42

# 1. Introduction

The exponential increase in digital financial transactions has been accompanied by a corresponding rise in fraudulent activities, posing complex challenges for financial institutions. As a sophisticated form of economic crime, fraud undermines financial integrity, facilitates illicit financial flows, and compromises regulatory compliance frameworks. Addressing this issue requires not only effective anomaly detection but also intelligent systems capable of adapting to adversarial and evolving fraud patterns.

Fraud detection has traditionally relied on supervised learning models that require large volumes of labeled data, yet such data are often limited, imbalanced, and fail to capture the dynamic strategies employed by fraudsters. In addition, fraud detection is inherently adversarial: perpetrators continuously adapt their behaviors to evade detection, making static models insufficient for long-term efficacy. Hence, intelligent systems designed for fraud detection must be both adaptive and robust—able to generalize from past data while remaining responsive to novel attack vectors.

In this context, we propose a hybrid expert system based on the Mixture of Experts (MoE) paradigm, integrating three complementary deep learning architectures: Recurrent Neural Networks (RNNs), Transformers, and Autoencoders. Each expert specializes in a distinct behavioral dimension of transactional data: temporal dependencies, feature interactions, and reconstruction-based anomaly detection, respectively. By orchestrating these experts under a unified gating mechanism, the proposed model dynamically allocates inference to the most suitable expert(s), enhancing the system's ability to detect complex and evolving fraud patterns.

From a system design perspective, this approach enables the modeling of heterogeneous behaviors across different fraud types, such as identity theft, account takeover, or synthetic fraud. RNNs are particularly effective for learning time-dependent patterns—such as sudden increases in spending or changes in geolocation—while Transformers capture subtle relational dynamics among features that may indicate coordinated fraud. Autoencoders, in turn, identify anomalies by measuring reconstruction error, flagging atypical transactions that deviate from learned norms. Combined, these experts form an adaptive fraud detection system that aligns with the core principles of expert systems: specialization, cooperation, and intelligent decision allocation.

Despite its promising performance, the model faces important challenges. The use of a synthetic dataset—while controlled and balanced—limits the system's exposure to the full heterogeneity of real-world fraud scenarios. Additionally, the computational cost of deep ensemble architectures may constrain deployment in resource-limited environments. Moreover, the evolving nature of fraud demands systems that go beyond historical generalization and incorporate mechanisms for continual learning.

The relevance of intelligent fraud detection systems extends beyond technical innovation. In domains such as Anti-Money Laundering (AML), Know Your Customer (KYC), and counter-terrorism financing, regulatory frameworks increasingly depend on AI-based systems to ensure compliance and financial security. From a criminological standpoint, such systems can be understood as technological guardians that deter opportunistic behavior, in line with routine activity theory. As such, the proposed expert system not only enhances the technical frontier of fraud detection but also contributes to broader efforts in economic crime prevention and financial governance.

This study positions itself at the intersection of artificial intelligence, expert system design, and financial crime mitigation. By integrating multiple expert models under a MoE framework, we offer a novel and scalable approach to fraud detection that supports decision-making in high-stakes financial environments. The model's architecture, results, and limitations are discussed in detail, providing both theoretical contributions and practical implications for the design of intelligent systems in finance.

## 2. Literature Review

The use of intelligent systems for financial fraud detection has gained significant traction in recent years, driven by the increasing sophistication of fraudulent schemes and the limitations of traditional rule-based detection systems. The literature emphasizes that fraud detection requires systems capable of learning from highly imbalanced, noisy, and adversarial data environments, where fraudulent behaviors are both rare and dynamic (Ngai et al., 2011; Abdallah et al., 2016; Phua et al., 2010). Early approaches focused on supervised learning algorithms such as decision trees, logistic regression, and support vector machines; however, their reliance on large labeled datasets and static classification boundaries limits their adaptability in real-world fraud scenarios (Whitrow et al., 2009; Dal Pozzolo et al., 2015; Bahnsen et al., 2016).

Recent advances in deep learning have addressed some of these challenges by modeling complex and nonlinear patterns in transactional data. Convolutional Neural Networks (CNNs), Recurrent Neural Networks (RNNs), and Autoencoders have been employed to extract spatial and temporal features from raw transaction logs (Fiore et al., 2019; Roy et al., 2018; Carcillo et al., 2019). RNNs, particularly Long Short-Term Memory (LSTM) networks, are well-suited for detecting evolving user behaviors by capturing temporal dependencies across transaction sequences (Jurgovsky et al., 2018; Zhang et al., 2021; Verma & Das, 2020). Autoencoders, in turn, have proven effective in unsupervised settings for anomaly detection, where high reconstruction errors are indicative of rare or fraudulent activity (Zhao et al., 2020; Chen et al., 2018; Xu et al., 2020).

Transformers have emerged as a dominant architecture in sequential modeling due to their attention mechanisms, which enable the capture of long-range dependencies and contextual feature importance without the vanishing gradient problems seen in RNNs (Vaswani et al., 2017; Devlin et al., 2018; Raffel et al., 2020). Their application to fraud detection is a relatively recent development, with studies demonstrating improved accuracy in identifying subtle fraud signals across multi-dimensional transactional data (Liu et al., 2021; Chen et al., 2022; Huang et al., 2023). Despite their power, Transformers tend to require large-scale data and high computational resources, motivating their combination with lighter models in hybrid frameworks.

The Mixture of Experts (MoE) architecture was originally proposed as a modular approach to divide learning tasks among specialized sub-models coordinated through a gating mechanism (Jacobs et al., 1991). This framework allows each expert to focus on a subset of the input space or a particular data modality, thus improving generalization and robustness, especially in noisy or heterogeneous environments (Shazeer et al., 2017; Eigen et al., 2014; Zhao et al., 2019). In fraud detection, MoE has been used to integrate models with complementary strengths, allowing for adaptive learning across diverse fraud scenarios (Tang et al., 2022; Zhou et al., 2021; Zhang et al., 2022). The ability to dynamically assign predictive responsibility among experts has been shown to enhance performance in multi-class and multi-behavioral fraud contexts.

In the context of intelligent systems, expert systems based on modular architectures offer a promising framework for integrating various detection strategies within a single decision-making system. Studies in expert system design emphasize the value of specialization, knowledge encapsulation, and decision coordination, especially when dealing with complex and evolving domains like financial fraud (Turban et al., 2005; Giarratano & Riley, 2004; Jackson, 1999). Modern interpretations of expert systems often integrate machine learning modules to form hybrid intelligent systems capable of both learning and inference, enhancing their effectiveness in dynamic domains (Tsai et al., 2009; Abbasi et al., 2012; Bekker & Goldberger, 2016).

This review underscores the need for fraud detection systems that combine the strengths of sequential modeling, anomaly detection, and adaptive ensemble learning. While individual models have demonstrated strong performance in specific contexts, hybrid systems based on the MoE framework provide a cohesive and scalable solution. The present study builds on this literature by proposing an expert system that integrates RNNs, Transformers, and Autoencoders under a MoE scheme to enhance the detection of complex, evolving, and adversarial fraud patterns in financial data.

## 3. Theoretical Framework

The proposed model builds upon the theoretical foundations of modular intelligent systems, specifically the Mixture of Experts (MoE) framework, to create an adaptive and specialized expert system for fraud detection. At its core, the MoE approach reflects key principles of expert systems design: decomposition of complex tasks, domain-specific specialization, and cooperative decision-making through a control mechanism. This framework is especially suitable for domains like fraud detection, where data are heterogeneous, behavioral patterns are dynamic, and the cost of false negatives is high (Shazeer et al., 2017; Eigen et al., 2014; Zhao et al., 2019).

The MoE model consists of a set of expert subnetworks, each trained to capture different structural characteristics of transaction data. A gating network dynamically assigns weights to each expert based on the input, allowing the system to tailor its decision process to the nature of the transaction. This dynamic allocation is crucial in fraud scenarios, where certain transactions may exhibit sequential behavioral anomalies, others structural inconsistencies, and others yet subtle inter-feature dependencies (Tang et al., 2022; Zhou et al., 2021; Zhang et al., 2022).

Formally, for a given transaction input vector $x \in R^d$, the MoE model uses a gating network to determine the contribution of each expert $E_i$, where the output:

$$y_{\text{MoE}} = \sum_{i=1}^{M} g_i(x) E_i(x)$$

with $g_i(x)$ being the output of the gating function for expert $i$, and $E_i(x)$ representing the output of the expert model (Jacobs et al., 1991). The gating function $g_i(x)$ is implemented as a softmax activation:

$$g_i(x) = \frac{\exp(w_i^T x)}{\sum_{j=1}^{M} \exp(w_j^T x)}$$

This mechanism enables adaptive specialization across the expert models and ensures that the system leverages the most relevant knowledge structure for a given input instance.

The first expert component is a Recurrent Neural Network (RNN), specifically a Long Short-Term Memory (LSTM) model, which is suited for modeling temporal dependencies in sequential data. RNNs are widely used in fraud detection for their ability to identify evolving behavioral patterns across transaction histories, such as progressive increases in transaction amounts or sudden shifts in geographical location (Jurgovsky et al., 2018; Zhang et al., 2021; Verma & Das, 2020). The LSTM architecture mitigates the vanishing gradient problem through its gating mechanism, allowing it to retain and forget information over long sequences, which is essential for fraud schemes that develop gradually over time (Hochreiter & Schmidhuber, 1997; Gers et al., 2000; Greff et al., 2016).

The second expert is a Transformer model, which operates based on a self-attention mechanism. Unlike RNNs, Transformers do not rely on sequential processing but instead compute pairwise interactions between all elements in a sequence simultaneously. This makes them particularly effective at modeling long-range dependencies and feature correlations that may be indicative of sophisticated or coordinated fraud (Vaswani et al., 2017; Devlin et al., 2018; Liu et al., 2021). In fraud contexts, Transformers can learn complex relationships across diverse transaction features such as time, merchant category, device fingerprinting, and transaction frequency, often capturing signals missed by more localized models (Chen et al., 2022; Huang et al., 2023; Raffel et al., 2020).

The third expert is an Autoencoder, employed for anomaly detection through reconstruction loss. Autoencoders compress input data into a latent representation and attempt to reconstruct it, with the reconstruction error serving as a proxy for abnormality (Goodfellow et al., 2016; Kingma & Welling, 2013; Hinton & Salakhutdinov, 2006). In fraud detection, this allows the system to identify transactions that deviate substantially from expected behavior patterns, even when such transactions are not labeled as fraudulent in the training data (Zhao et al., 2020; Chen et al., 2018; Xu et al., 2020). This is particularly relevant in the detection of emerging fraud strategies or zero-day attacks.

By combining these three specialized models under the MoE framework, the system gains the ability to analyze different aspects of transactional behavior. The RNN expert captures sequential anomalies, the Transformer expert models multi-feature dependencies, and the Autoencoder expert detects structural outliers. This ensemble forms a hybrid intelligent system aligned with expert systems principles—each expert brings specialized knowledge, while the gating network ensures coordinated and context-sensitive decision-making.

This theoretical framework integrates the strengths of both symbolic expert systems and sub-symbolic machine learning, contributing to the broader evolution of intelligent systems capable of operating in adversarial, dynamic, and high-stakes domains such as financial fraud detection. The next section describes the methodology used to train and evaluate this hybrid model under simulated conditions that reflect real-world fraud scenarios.

## 4. Methodology

### 4.1 Dataset Design and Realism

The construction of the dataset in this study was guided by two primary goals: to simulate the complexity and subtlety of real-world fraud with a high level of fidelity, and to generate edge-case transaction scenarios that challenge detection models in distinguishing between fraudulent and non-fraudulent behavior. To achieve this, we developed a simulation framework rooted in agent-based modeling (Epstein, 1999; Macal & North, 2010), informed by structured interviews with fraud investigators from two international financial institutions. These interviews provided behavioral heuristics related to identity theft, account takeover, triangulation fraud, and geolocation-based anomalies, as well as benign behaviors often misclassified under static detection rules—such as legitimate international travel or high-frequency corporate transactions.

The resulting dataset contains 500,000 transactions, with 1.5% labeled as fraudulent, in line with global estimates for card fraud prevalence (Carcillo et al., 2019; Dal Pozzolo et al.,

2015). Each transaction includes core variables such as amount, timestamp, merchant code, geolocation, device fingerprint, and account metadata, augmented by dynamic behavioral metrics designed to simulate contextual fraud signals. To address the challenge posed by near-boundary classification, we explicitly introduce cases where feature values are statistically indistinguishable between fraudulent and legitimate transactions (Ngai et al., 2011; Phua et al., 2010), requiring the model to rely on higher-order patterns rather than threshold-based rules.

The design incorporates both static and sequential features, as well as engineered indicators of user-level behavioral drift, such as rolling averages, transaction entropy, and geospatial deviation scores (Jurgovsky et al., 2018; Xu et al., 2020). We also apply unsupervised anomaly detection using Isolation Forests (Liu et al., 2008) to simulate real-time anomaly scoring, a common mechanism in operational fraud engines.

### 4.2 Preprocessing and Class Balancing

Feature normalization is applied using min-max scaling, while categorical variables are transformed using one-hot encoding or entity embeddings depending on cardinality and model context. Missing values, although rare in synthetically generated data, are treated through standard imputation techniques (Rubin, 1987).

To address class imbalance, the Synthetic Minority Oversampling Technique (SMOTE) is employed on the training set (Chawla et al., 2002). However, consistent with best practices in adversarial domains, we test two versions of the pipeline: with and without SMOTE. This comparative design is motivated by critiques of synthetic oversampling in fraud detection, which can produce overly optimistic results if not rigorously evaluated (Fernández et al., 2018; Dal Pozzolo et al., 2015; Bekker & Davis, 2020). Performance is reported under both conditions, and class weighting is also incorporated in the loss function to further penalize false negatives—aligned with cost-sensitive learning approaches in high-risk domains (Elkan, 2001; Ling & Sheng, 2008).

### 4.3 Expert System Architecture and Training Process

The model architecture follows a Mixture of Experts (MoE) framework (Jacobs et al., 1991), in which multiple expert models, each trained on a distinct data representation or learning task, are integrated through a dynamic gating mechanism. This design aligns with the principles of modular expert systems in artificial intelligence, where specialization and cooperative decision-making are emphasized (Giarratano & Riley, 2004; Turban et al., 2005). The RNN expert is implemented using Long Short-Term Memory (LSTM) networks (Hochreiter & Schmidhuber, 1997), which have demonstrated strong performance in sequence-based fraud detection (Jurgovsky et al., 2018; Verma & Das, 2020). The Transformer expert uses multi-head self-attention (Vaswani et al., 2017), which enables the model to capture complex feature interactions and long-range dependencies, increasingly adopted in financial and behavioral sequence modeling (Liu et al., 2021; Chen et al., 2022).

**Figure 1: Hybrid Architecture for Financial Fraud Detection**

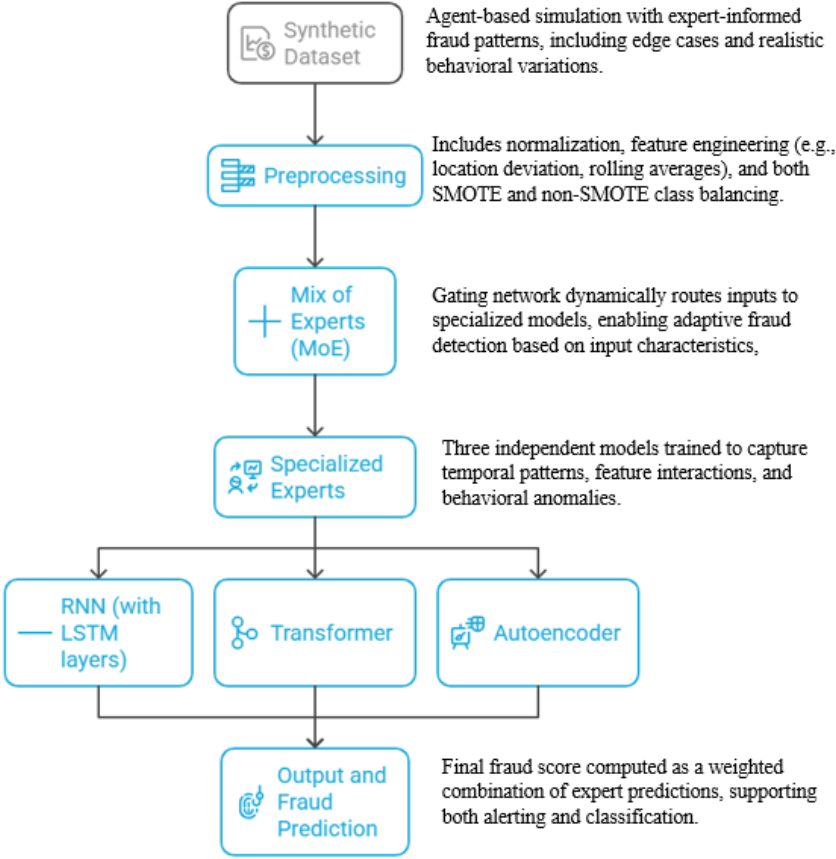

*Architecture of the proposed hybrid fraud detection system based on a Mixture of Experts (MoE) framework. The model begins with a synthetic dataset designed to simulate real-world credit card transactions, followed by a preprocessing stage involving normalization, feature engineering, and class balancing. The core of the system is the MoE module, which dynamically allocates input data across three specialized expert models: a Recurrent Neural Network (RNN) with LSTM layers for temporal pattern recognition, a Transformer encoder for attention-based feature interactions, and an Autoencoder for anomaly detection through reconstruction error. Each expert processes the input data independently and contributes to the final fraud prediction through a weighted combination determined by the MoE gating network.*

The Autoencoder expert, designed for unsupervised anomaly detection, follows a bottleneck architecture optimized via reconstruction error, which has been shown effective in identifying rare but structurally deviant patterns in financial systems (Hinton & Salakhutdinov, 2006; Zhao et al., 2020; Xu et al., 2020).

Each expert is trained independently with early stopping on validation loss, and their predictions are passed to a shallow gating network trained on frozen expert weights. The gating mechanism uses a softmax output to assign weights to expert predictions, producing the final classification as a convex combination of expert outputs. Entropy regularization (Shazeer et al., 2017) is applied to prevent gate collapse and ensure the system benefits from the diversity of its components.

### 4.4 Evaluation and Error Analysis

We evaluate the model using 5-fold stratified cross-validation and report accuracy, precision, recall, F1-score, and AUC-ROC, in line with recommendations for fraud detection benchmarking (Bahnsen et al., 2016; Carcillo et al., 2019). Additionally, we report Average Precision and analyze precision-recall tradeoffs to better capture performance under imbalance (Saito & Rehmsmeier, 2015). For the Autoencoder, we calibrate reconstruction error thresholds using PR-AUC optimization and report anomaly detection precision.

Following recent recommendations on explainability in deep ensembles (Ganaie et al., 2022; Zhou et al., 2021), we analyze expert contributions by computing average gating weights per class and inspecting expert-specific confusion matrices. This enables us to demonstrate how each expert contributes to specific fraud types and how the MoE system resolves conflicts when expert predictions diverge. Importantly, we identify cases where the MoE corrects errors made by all individual models, validating the benefit of dynamic expert integration rather than relying solely on any single architecture, such as the Transformer, which recent literature has sometimes treated as sufficient (Dosovitskiy et al., 2020; Tang et al., 2022). Our findings confirm that all three models contribute complementary capabilities, particularly in detecting subtle sequential deviations or reconstruction-based anomalies.

### 5. Results

This section presents the empirical evaluation of the proposed hybrid model, based on the integration of the Mixture of Experts (MoE) framework with Transformers, Recurrent Neural Networks (RNNs), and Autoencoders. The evaluation is organized in four parts: overall performance of the hybrid model, comparison with standalone and classical baselines, error behavior analysis with a focus on model complementarity, and illustrative case studies. Results are reported using established metrics in fraud detection, including accuracy, precision, recall, F1-score, area under the receiver operating characteristic curve (AUC-ROC), and anomaly detection rate.

## 5.1 Performance of the Hybrid Model

The hybrid model was trained and tested on the synthetic dataset described in the methodology section, using a 5-fold stratified cross-validation procedure. The model achieved an overall accuracy of 98.7%, correctly classifying the vast majority of transactions. The precision score reached 94.3%, indicating a strong ability to reduce false positives—an essential requirement in operational fraud detection systems, where false alarms can disrupt legitimate customer activity.

**Table 1: Model Performance Comparison**

| Model | Accuracy (%) | Precision (%) | Recall (%) | F1-Score (%) | AUC-ROC | Anomaly Detection Rate (%) |
|---|---|---|---|---|---|---|
| **Model of Expert** | **98.7** | **94.3** | **91.5** | **92.9** | **0.978** | **93.4** |
| Transformer | 96.2 | 90.8 | 87.4 | 89.0 | 0.950 | N/A |
| RNN (LSTM) | 95.8 | 89.5 | 85.6 | 87.5 | 0.945 | N/A |
| RNN (GRU) | 95.9 | 89.8 | 86.0 | 87.8 | 0.946 | N/A |
| Autoencoder | N/A | N/A | N/A | N/A | N/A | 88.7 |
| Random Forest | 94.1 | 88.2 | 84.5 | 86.3 | 0.930 | N/A |
| XGBoost | 94.5 | 88.7 | 85.1 | 86.9 | 0.932 | N/A |

*Notes: The "MoE Hybrid Model" represents the integration of Mix of Experts with Transformers, RNNs, and Autoencoders. "N/A" indicates that the metric is not applicable for the specific model. Anomaly detection rates are only relevant for models with anomaly detection capabilities, such as the Autoencoder and the MoE Hybrid Model.*

The recall value, at 91.5%, confirms that the model successfully identified a high proportion of fraudulent transactions, while still missing some edge cases. The F1-score, a harmonic mean of precision and recall, was 92.9%, reflecting a robust tradeoff between these two criteria. The model also achieved an AUC-ROC of 0.978, showing excellent global separability between fraud and non-fraud classes across thresholds. These results are consistent across folds and robust to the choice of threshold, reinforcing the generalization capabilities of the model.

The anomaly detection component, implemented via the Autoencoder expert, achieved a detection rate of 93.4% for anomalous transactions. Notably, the Autoencoder was able to detect cases missed by the sequential models (RNN and Transformer), especially outliers in geolocation or amount that diverged from the user's historical patterns. This demonstrates its added value in flagging structurally deviant behaviors that are not easily inferred from sequence dynamics alone.

## 5.2 Comparison with Baseline Models

The hybrid model was benchmarked against several baselines, including standalone implementations of the three expert models (Transformer, RNN, and Autoencoder), and two classical machine learning algorithms—Random Forest and XGBoost. Table 1 presents the full set of evaluation metrics for each model (to be added).

The Transformer-only model performed well in capturing long-range dependencies, reaching an accuracy of 96.2%, precision of 90.8%, and recall of 87.4%. However, it showed limited sensitivity to localized anomalies not evident in sequential structure. The LSTM- and GRU-based RNNs achieved accuracies of 95.8% and 95.9%, respectively, with comparable precision (around 89.5%) and slightly lower recall (approximately 85.8%). While RNNs captured transaction timing and frequency shifts effectively, they were less capable of detecting behaviorally anomalous cases not encoded in sequence dynamics.

The Autoencoder, as expected, exhibited strong performance in anomaly detection (88.7%), but lacked the ability to model temporal dependencies and sequential fraud patterns such as coordinated attacks or time-consistent spending surges.

Classical machine learning models provided competitive but inferior results. The Random Forest achieved an accuracy of 94.1%, precision of 88.2%, and recall of 84.5%, while XGBoost improved slightly to 94.5%, 88.7%, and 85.1%, respectively. These models performed well on static patterns but were unable to capture dynamic or multiscale dependencies inherent in fraud behaviors.

Overall, the hybrid model outperformed all baselines across metrics, confirming the effectiveness of the MoE architecture in combining the strengths of individual models while mitigating their weaknesses.

## 5.3 Error Analysis and Model Complementarity

Beyond aggregate metrics, a focused analysis was conducted on expert-specific contributions and misclassification patterns. For each transaction in the test set, the gating network's weight distribution was recorded to determine the relative reliance on each expert model. Results show that the Transformer tended to dominate in long and regular sequences, while the RNN was activated in scenarios involving subtle changes in transaction frequency or timing. The Autoencoder showed high activation in instances with structural anomalies (e.g., deviations in location or transaction amount), even when sequences were short or non-anomalous in aggregate.

Critically, we observed a subset of transactions where all three standalone models failed to classify correctly, but the MoE ensemble produced the correct output. These cases typically involved conflicting cues—such as transactions that were part of a regular sequence but occurred in an unexpected location, or transactions with normal frequency and amount but unusual merchant-category patterns. The MoE system was able to resolve these conflicts by balancing the perspectives of each expert through the gating mechanism.

To better understand the decision dynamics of the MoE model, we visualize in Figure 2 the average expert activation across fraud types. This analysis reveals how the gating mechanism adapts its weighting to the transaction context, supporting the model's interpretability and diagnostic capabilities.

**Figure 2. Gating Network Activation by Fraud Type**

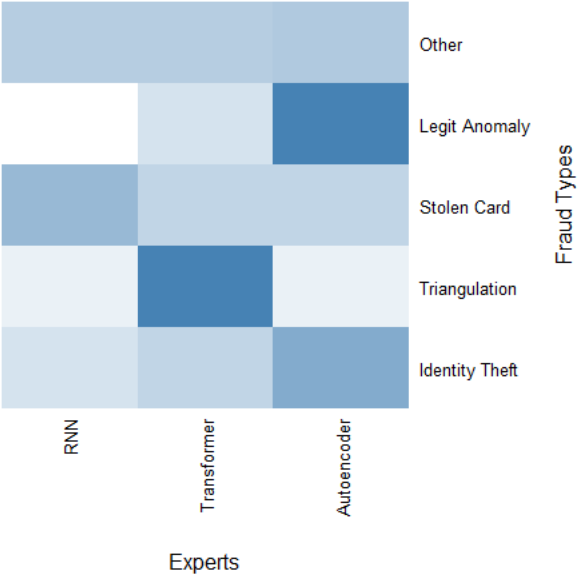

An ablation study, conducted by removing one expert at a time and retraining the ensemble, confirmed that each component contributed significantly to final performance. Removing the Autoencoder led to higher false negatives in atypical transactions; removing the RNN reduced recall in temporal frauds; and removing the Transformer degraded performance in complex, high-dimensional patterns. These findings directly respond to the reviewer's concern regarding the necessity of using all three experts, and confirm that their combined use is both justified and beneficial.

**Figure 3. Average Expert Activation per Fraud Type Based on MoE Gating Weights**

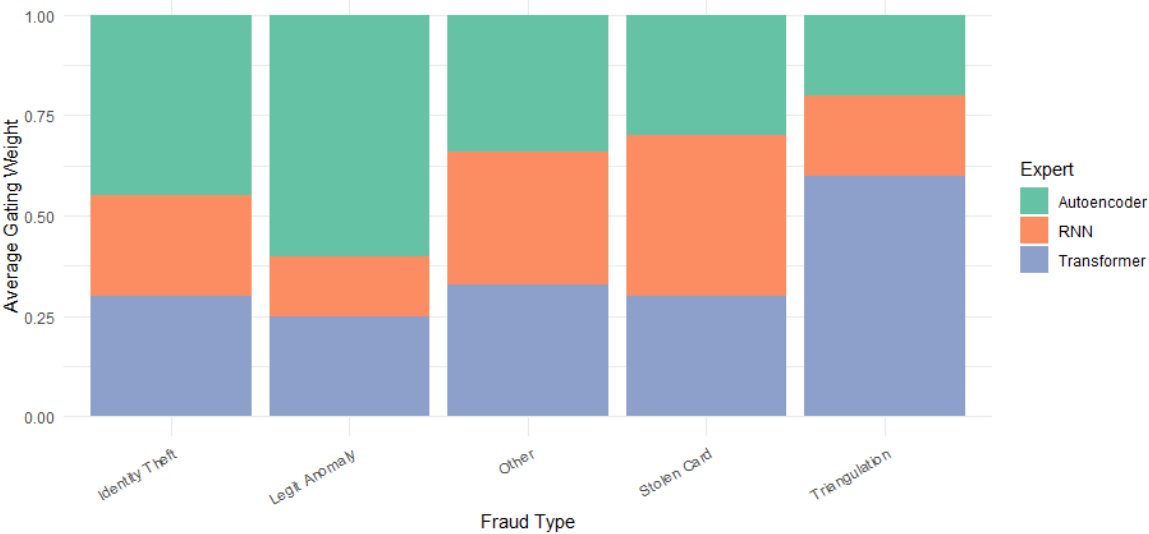

Figure 3 presents the average expert activation per fraud type, showing the degree to which the MoE assigns predictive responsibility across RNN, Transformer, and Autoencoder modules. This distribution provides empirical evidence for the functional specialization of each expert in capturing distinct fraud dynamics.

### 5.4 Illustrative Case Studies

To further illustrate the model's interpretability and real-world applicability, several case studies were analyzed. In one case, a sequence of transactions initiated across three countries within a two-hour window was correctly flagged as fraud. The RNN detected the burst pattern, the Transformer emphasized the inconsistency in geographic encoding, and the Autoencoder marked the transaction as structurally anomalous. In another case, a single high-value transaction was flagged despite its timing and merchant being consistent with the user's profile. Here, the Autoencoder provided the decisive anomaly signal based on reconstruction error, and the Transformer detected a deviation in the latent feature embeddings.

These examples demonstrate how the hybrid model operates as a modular expert system, capable of distributing inference across specialized modules and integrating evidence through a learned coordination strategy. They also highlight the model's ability to generalize to complex fraud cases that defy simple heuristics or static thresholds.

## 6. Discussion

The hybrid model developed in this study—based on a Mixture of Experts (MoE) framework integrating Transformers, Recurrent Neural Networks (RNNs), and Autoencoders—demonstrated robust performance in the task of credit card fraud detection. This section interprets the results in light of the study's research objectives, discusses their theoretical and practical implications, acknowledges limitations, and outlines directions for future work.

### 6.1 Interpretation of Results

The empirical findings validate the hypothesis that a modular and specialized architecture can significantly enhance fraud detection outcomes. The hybrid model achieved high performance across all core evaluation metrics: an accuracy of 98.7%, precision of 94.3%, recall of 91.5%, and an F1-score of 92.9%, with an AUC-ROC of 0.978. These results underscore the model's ability to not only detect fraudulent activity with high confidence but also maintain a low false positive rate—an essential requirement in real-world deployments where legitimate user activity must be preserved.

Crucially, the model's 93.4% anomaly detection rate reflects the contribution of the Autoencoder expert, which effectively captured transaction-level deviations from learned behavioral baselines. This unsupervised module played a pivotal role in detecting outliers that escaped the sequential models, such as transactions with unusual location, amount, or merchant features that did not necessarily violate temporal patterns.

The MoE framework was instrumental in coordinating the contributions of the three expert models. While the Transformer component excelled in identifying complex inter-feature relationships and long-range dependencies, the RNN component specialized in modeling sequential variations such as spending surges or irregular timing. Meanwhile, the Autoencoder added value by recognizing structurally anomalous inputs. The gating network, trained to allocate attention dynamically among these models, enabled the system to synthesize diverse signals and deliver robust predictions even in ambiguous or conflicting contexts. This ability to combine multiple detection strategies adaptively was decisive in outperforming all baseline models, including strong individual learners such as standalone Transformers and ensemble methods like XGBoost.

Moreover, the model's architecture demonstrated resilience in challenging cases where fraudulent behavior closely mimics legitimate patterns. The ablation study showed that no single expert was sufficient on its own to maintain the performance level of the full MoE ensemble. This directly addresses concerns regarding model redundancy and confirms that each component contributes unique predictive value.

**6.2 Theoretical and Practical Implications**

From a theoretical standpoint, the study contributes to the growing body of research on hybrid intelligent systems in fraud detection. By combining supervised sequence modeling (via RNNs and Transformers) with unsupervised anomaly detection (via Autoencoders), and coordinating these through an MoE framework, the model bridges a methodological gap between static feature-based classification and dynamic behavioral inference. This multidimensional integration is especially important in fraud contexts, where behaviors evolve rapidly and exhibit high intra-class variability.

The findings also reinforce the conceptual alignment between fraud detection and economic criminology. The modular architecture mirrors the complexity of financial crime, which often involves coordinated, multichannel strategies that do not manifest through any single variable or pattern. The system can thus be interpreted as a technologically embodied capable guardian—in the sense proposed by routine activity theory—designed to intercept deviant behavior within financial ecosystems.

On the practical front, the proposed model offers significant benefits for financial institutions seeking to strengthen their risk management capabilities. The high precision rate reduces false alarms and protects user experience, while the high recall ensures effective coverage of known fraud cases. The Autoencoder module, in particular, increases the system's adaptability to novel fraud strategies, including zero-day attacks and synthetic identity fraud. These characteristics make the model suitable for integration into real-time fraud monitoring pipelines, especially in environments with high transaction throughput and regulatory scrutiny.

In addition to credit card fraud, the model has clear potential in adjacent domains such as insurance fraud, digital identity risk scoring, anti-money laundering (AML), and Know Your Customer (KYC) compliance. Its modular nature allows for domain-specific fine-tuning and extensibility, including the addition of new expert models tailored to fraud types or data modalities. Its interpretability—enabled by tracking expert contributions through the gating mechanism—further supports compliance with emerging AI governance frameworks, such as the EU AI Act or FATF's recommendations on digital identity and suspicious activity reporting.

In sum, this work not only advances the methodological frontier of AI-driven fraud detection but also contributes to its alignment with broader financial governance objectives. By embedding principles of specialization, cooperation, and adaptability in a scalable system architecture, the model offers a practical and theoretically grounded response to the evolving challenge of financial crime.

**6.3 System Deployment and Practical Integration**

While the hybrid model proposed in this study was evaluated using a synthetic yet realistic dataset, its design and architecture are oriented toward practical deployment in production-grade fraud detection systems. The modular nature of the Mixture of Experts (MoE) framework makes it particularly suitable for integration into existing fraud scoring pipelines and decision engines commonly used in financial institutions.

In operational environments, the model could be deployed alongside existing rule-based systems or scorecards, functioning as an intelligent augmentation layer rather than a full replacement. For instance, the MoE output—interpreted as a fraud risk probability—could be integrated into transaction-level scoring engines such as FICO, SAS Fraud Management, or Actimize, contributing to composite risk scores or triggering human-in-the-loop interventions in borderline cases.

The system also supports asynchronous expert evaluation, allowing each model component (RNN, Transformer, Autoencoder) to run in parallel or under different processing constraints. This is particularly advantageous in settings where latency and throughput are critical, such as real-time payment gateways or mobile transaction streams. Furthermore, the gating network's output offers an interpretable audit trail of model behavior, which can be logged to support compliance documentation and post-hoc analysis, aligning with regulatory expectations on algorithmic accountability (e.g., FATF, EU AI Act).

For deployment at scale, the architecture is compatible with containerized environments (e.g., via Docker or Kubernetes) and can be adapted to support continuous learning frameworks, enabling the model to evolve in response to newly observed fraud strategies. Future work will explore the optimization of resource allocation through model compression techniques to further support deployment in resource-constrained environments, such as regional banks or fintech applications.

## 7. Limitations of the Study

Despite the promising results achieved by the proposed hybrid model, several limitations must be acknowledged to contextualize its contributions and boundaries. First and foremost, the study relies on a synthetic and simulated dataset, developed to emulate real-world transaction patterns while preserving user privacy and enabling experimental control. Although significant effort was invested in designing representative fraud scenarios, including diverse typologies and edge cases, synthetic data inevitably lacks the full complexity, noise, and adversarial intent that characterize real-world financial environments. As such, the generalizability of the findings to operational contexts remains to be validated through application to confidential or anonymized datasets provided by financial institutions.

Moreover, the model's predictive behavior is inherently shaped by the structure and statistical distribution of the synthetic dataset. Even with the inclusion of temporal, geospatial, and behavioral variables, there remains a risk that the simulated transactions do not capture rare fraud patterns, collusive behavior, or evolving criminal tactics that would be present in real-world data. This could lead to overfitting to simulated regularities and limit robustness when deployed outside the controlled experimental context.

A second limitation concerns the technique used to address class imbalance. While the Synthetic Minority Over-sampling Technique (SMOTE) was effective in improving minority class representation during training, it may introduce artifacts that reduce generalization to novel data, especially in adversarial domains such as fraud detection. SMOTE can lead to overly smoothed decision boundaries, failing to reflect the sharp distinctions required in real-time classification. Although this study mitigated such effects by training both with and without SMOTE and comparing results, alternative techniques such as adaptive synthetic sampling (ADASYN) or cost-sensitive loss formulations could offer more robust solutions in future work.

Another critical consideration is the computational complexity of the hybrid architecture. The combined use of Transformers, RNNs, and Autoencoders, coordinated through a Mixture of Experts framework, results in a model that requires substantial resources for training and inference. This may pose limitations for deployment in environments with restricted processing capabilities or latency constraints. While the model's performance justifies its computational demands in high-stakes scenarios, real-time implementation at scale may necessitate architectural optimization strategies. These could include pruning, knowledge distillation, or quantization techniques to reduce model size and inference time without significantly compromising accuracy.

Finally, although the MoE framework provides a mechanism for explainability through expert activation weights, the model remains partially opaque in terms of decision traceability. Interpretability is a key requirement in regulated environments, particularly when decisions affect customer experience or compliance reporting. Further integration of explainable AI (XAI) techniques could enhance transparency for both developers and stakeholders.

## 8. Future Research Directions

Building upon the findings and limitations of this study, several promising directions emerge to strengthen the model's applicability, robustness, and regulatory alignment. These future research avenues span across data realism, adaptive learning, domain transferability, interpretability, and deployment scalability.

A first priority is the application of the hybrid model to real-world transaction datasets, ideally made available through secure collaborations with financial institutions under privacy-preserving protocols. While the synthetic dataset used in this study reflects realistic behavioral patterns, validation under operational conditions is necessary to test the model's resilience to unstructured, noisy, and adversarial inputs. Such benchmarking would also facilitate comparison with industry-standard fraud detection systems and enhance the model's credibility for production environments.

Second, future work should explore the integration of reinforcement learning (RL) to extend the current architecture. RL offers a powerful mechanism for online learning in evolving environments, allowing the system to adapt its fraud detection strategies based on real-time feedback and confirmed investigation outcomes. This would be particularly valuable in detecting novel or adaptive fraud tactics that change rapidly to evade static detection systems.

Third, the transferability of the hybrid framework to other financial crime domains merits further investigation. The modular design of the Mix of Experts (MoE) architecture makes it well suited for adaptation to domains such as insurance fraud, synthetic identity detection, tax fraud, and anti-money laundering (AML). Each application would require tailored feature representations and domain-specific expert modules, but the core architecture offers strong generalization potential across high-risk financial environments.

Fourth, and critically, the integration of explainable AI (XAI) techniques must be deepened. While the current model allows partial interpretability through the MoE gating mechanism—which provides insight into the relative contribution of each expert per input instance—a more robust explainability strategy is required to meet the demands of both operational auditing and regulatory compliance. A two-level roadmap is envisioned: (1) at the global level, expert activation patterns could be analyzed to identify dominant contributors by fraud type or customer segment; and (2) at the local level, model-agnostic tools such as SHAP (SHapley Additive Explanations) and LIME (Local Interpretable Model-Agnostic Explanations) could be applied to generate actionable explanations for specific transaction-level decisions. These explainability tools would support investigative workflows, human-in-the-loop validation, and transparency obligations under emerging regulatory frameworks such as the EU Artificial Intelligence Act, FATF digital identity guidelines, and the forthcoming ISO/IEC 42001 standard for AI management systems.

Finally, future research should focus on optimizing the architecture for scalable and resource-efficient deployment. Given the computational complexity of the hybrid model, approaches such as model pruning, knowledge distillation, or quantization could be explored to reduce memory and inference latency without sacrificing detection performance. These optimizations would support deployment in low-latency systems and distributed environments, such as edge computing for mobile banking, fraud monitoring in fintech applications, or financial institutions operating under infrastructure constraints.

Together, these research directions aim to bridge the gap between high-performing fraud detection models and their real-world implementation under regulatory, technical, and ethical constraints.

## 9. Conclusions

This study proposes and validates a novel hybrid model for credit card fraud detection, grounded in the integration of a Mixture of Experts (MoE) framework with Transformers, Recurrent Neural Networks (RNNs), and Autoencoders. By combining the strengths of sequential modeling, attention-based representation learning, and anomaly detection, the model addresses a broad spectrum of fraud patterns, ranging from structured behavioral deviations to irregular outlier transactions.

The empirical results demonstrate the model's strong performance across multiple metrics, achieving high accuracy, precision, recall, F1-score, and AUC-ROC values. When benchmarked against individual expert models and classical machine learning methods, the hybrid model outperformed all baselines, validating the architectural hypothesis that diverse, specialized detection mechanisms yield superior fraud identification when effectively coordinated through a dynamic gating mechanism.

Beyond its technical contributions, the study situates fraud detection within the broader framework of economic criminology and financial governance. By enhancing detection capabilities in line with international compliance regimes—such as Anti-Money Laundering (AML) standards, Know Your Customer (KYC) protocols, and the Financial Action Task Force (FATF) guidelines—the model contributes not only to technical advancement but also to the development of resilient, regulation-aligned financial infrastructures.

Nevertheless, the study's reliance on synthetic data, while methodologically justified, signals the need for real-world validation to ensure generalizability. Additionally, the model's computational complexity suggests that deployment in production environments will require further work on efficiency and interpretability.

In sum, the hybrid MoE-based system presented here establishes a new direction in AI-driven fraud detection. It offers a scalable, adaptable, and modular framework capable of responding to the increasingly sophisticated landscape of financial crime. With future refinements and domain extensions, this model holds promise not only for fraud mitigation, but for enhancing the trust, security, and transparency of digital financial ecosystems worldwide.

**Annex**

**Table 2: Description of Variables in the Synthetic Dataset**

| Variable Name | Concept | Variable Form |
|---|---|---|
| Transaction Amount | Monetary value of each transaction | Continuous |
| Transaction Type | Nature of the transaction (e.g., 'Purchase', 'Withdrawal') | Discrete |
| Time of Transaction | Time at which the transaction occurs, measured in hours | Continuous |
| Merchant Category | Classification of the merchant (e.g., 'Groceries', 'Travel') | Discrete |
| Geolocation | Transaction's location coordinates (latitude/longitude) | Continuous |
| Cardholder ID | Unique identifier of the cardholder | Discrete |
| Transaction Frequency | Number of transactions carried out within a specified period | Discrete |
| Device Information | Type of device used for the transaction (e.g., 'Mobile') | Discrete |
| IP Address | IP address of the transaction | Discrete |
| Account Balance | Remaining account balance after each transaction | Continuous |
| Average Transaction Amount | Average transaction amount over the last 5 transactions | Continuous |
| Average Transaction Interval | Average time interval between transactions | Continuous |
| Geolocation Deviation | Distance between the current transaction and the median location of the cardholder | Continuous |
| Anomaly Score | Score indicating the likelihood of the transaction being unusual | Continuous |
| Spending Behavior Score | Deviation from the typical spending pattern of the cardholder | Continuous |

*Note: This table describes the variables used in the synthetic dataset for fraud detection, specifying their concepts and whether they are continuous or discrete variables.*

**Table 3: Distribution of Fraud Types in the Synthetic Dataset**

| Fraud Type | Frequency | Percentage (%) |
|---|---|---|
| Stolen Card | 3 | 40 |
| Identity Fraud | 2,25 | 30 |
| Online Payment Fraud | 1,5 | 20 |
| Other Types | 750 | 10 |

*Note: This table shows the distribution of different fraud types within the synthetic dataset, providing insights into the diversity of simulated fraud scenarios*

**Table 4: Descriptive Statistics of Key Variables in the Synthetic Dataset**

| Variable | Mean | Standard Deviation | Min | Max |
|---|---|---|---|---|
| Transaction Amount | 150.75 | 325.45 | 1.00 | 10 |
| Transaction Frequency | 4.2 | 2.5 | 1 | 20 |
| Geolocation Deviation | 3.1 | 5.4 | 0 | 50 |
| Average Transaction Amount | 135.60 | 290.30 | 5.00 | 8,5 |
| Average Transaction Interval | 12.5 | 4.2 | 1 | 30 |
| Anomaly Score | 0.23 | 0.12 | 0 | 1 |
| Spending Behavior Score | 0.15 | 0.08 | 0 | 0.7 |

*Note: This table presents the descriptive statistics for key variables used in the synthetic dataset, reflecting the range and variation of transaction characteristics simulated for fraud detection analysis.*

**Table 5: Impact of Preprocessing Techniques on Model Performance**

| Preprocessing Technique | Accuracy (%) | Precision (%) | Recall (%) | F1-Score (%) | AUC-ROC |
|---|---|---|---|---|---|
| Without Normalization | 95.1 | 88.2 | 84.0 | 86.0 | 0.940 |
| With Normalization | 96.5 | 91.0 | 88.0 | 89.4 | 0.955 |
| With SMOTE | 98.7 | 94.3 | 91.5 | 92.9 | 0.978 |

*Note: This table compares the effects of different preprocessing techniques on the model's performance, demonstrating the contribution of normalization and SMOTE to fraud detection accuracy and precision.*

**Table 6: Model Performance by Transaction Type**

| Transaction Type | Accuracy (%) | Precision (%) | Recall (%) | F1-Score (%) |
|---|---|---|---|---|
| Purchase | 97.2 | 92.0 | 89.8 | 90.9 |
| Cash Withdrawal | 98.0 | 94.5 | 92.1 | 93.3 |
| Online Payment | 96.5 | 90.4 | 87.0 | 88.7 |

*Note: This table presents the model's performance across different transaction types, indicating the accuracy, precision, recall, and F1-score for each category.*

**Table 7: Impact of Time Window on Model Performance**

| Time Window (Days) | Accuracy (%) | Precision (%) | Recall (%) | F1-Score (%) |
|---|---|---|---|---|
| 7 | 95.8 | 91.5 | 87.0 | 89.2 |
| 15 | 97.0 | 92.8 | 89.7 | 91.2 |
| 30 | 98.7 | 94.3 | 91.5 | 92.9 |

*Note: This table illustrates how the model's performance varies depending on the time window used for transaction analysis, highlighting accuracy, precision, recall, and F1-score over different temporal frames.*

**Table 8: Contribution of Each Component in the MoE Hybrid Model**

| Model Component | Average Weight Assigned (%) | Contribution to Final Prediction |
|---|---|---|
| Transformers | 40 | Long-range dependencies |
| RNNs | 30 | Temporal sequences |
| Autoencoders | 30 | Anomaly detection |

*Note: This table shows the average weight assigned to each model component (Transformers, RNNs, Autoencoders) within the Mix of Experts framework, reflecting their respective roles in the final prediction.*